\pdfoutput=1

\documentclass[prd,preprintnumbers,twocolumn,superscriptaddress,a4paper,showkeys,10pt,nofootinbib]{revtex4}

\usepackage{graphicx}
\usepackage{dcolumn}
\usepackage{bm}
\usepackage{multirow}
\usepackage{tabularx}
\usepackage{hyperref}
\usepackage{commath}
\usepackage{caption}
\usepackage{subcaption}
\usepackage{hyperref}
\hypersetup{colorlinks=true, citecolor=blue, urlcolor=blue, linkcolor=blue}

\begin{document}

\title{QCD probes at LHC}
\preprint{CMS-CR-2018/031}

%%%%%%%%%%%%%%%%%%%%%%%%%%%%%%%%%%%%%%%%%%%%%%%%%%%%%%%%%%%%%%%%%%

\author{G. Gil da Silveira}

\email[]{gustavo.silveira@cern.ch}

\affiliation{Instituto de F\'{\i}sica, Universidade Federal do Rio Grande do Sul\\
Caixa Postal 15051, CEP 91501-970, Porto Alegre, RS, Brazil}

\affiliation{Departamento de F\'{\i}sica Nuclear e de Altas Energias, Universidade do Estado do Rio de Janeiro\\
CEP 20550-013, Rio de Janeiro, RJ, Brazil}

\collaboration{on behalf of the ATLAS, CMS, and LHCb Collaborations}

%%%%%%%%%%%%%%%%%%%%%%%%%%%%%%%%%%%%%%%%%%%%%%%%%%%%%%%%%%%%%%%%%%

\begin{abstract}

The LHC experiments have reported new results with respect to the dynamics of the strong interactions in $pp$, $p$A, and AA collisions over the past years. In proton-proton collisions, the data analyses have focused in exploring the nature of underlying events and double parton scattering at high energies. For large systems, the heavy-ion collisions have provided new insights on physics aspects related to azimuthal correlations, jet quenching, and particle production, such as antiprotons. This Letter reports the recent results from the ATLAS, CMS, and LHCb Collaborations on these various topics and highlights its relevant findings for the high-energy community.

\end{abstract}

%%%%%%%%%%%%%%%%%%%%%%%%%%%%%%%%%%%%%%%%%%%%%%%%%%%%%%%%%%%%%%%%%%


\maketitle

%%%%%%%%%%%%%%%%%%%%%%%%%%%%%%%%%%%%%%%%%%%%%%%%%%%%%%%%%%%%%%%%%%

\section{Introduction}

During the past years, the experiments at the LHC have focused in studying several aspects of the strong interactions to provide high precision measurements. Apart of cross section measurements, the LHC experiments are performing a variety of studies to test the theoretical modelling in Monte Carlo (MC) event generators. These studies are mainly devoted to improve the theoretical description of physical phenomena with high-precision measurements, to explore production mechanisms, like multiple parton scattering (MPI), and to understand experimental signatures, as the jet quenching effect and azimuthal correlations in nuclear collisions. In the following, several results from the ATLAS, CMS, and LHCb Collaborations in proton-proton ($pp$), proton-Lead ($p$Pb), and Lead-Lead (PbPb) collisions are presented.

%%%%%%%%%%%%%%%%%%%%%%%%%%%%%%%%%%%%%%%%%%%%%%%%%%%%%%%%%%%%%%%%%%

\section{Proton-proton collisions}
\label{sec:pp}

%%%%%%%%%%%%%%%%%%%%%%%%%%%%%%%%%%%%%%%%%%%%%%%%%%%%%%%%%%%%%%%%%%

\subsection{Underlying events}
\label{sec:underl}

Most of the hadron-hadron collisions result in a high transferred momentum event, so-called {\it{hard scattering}}, where one parton from each colliding hadron contribute to produce a high transverse momentum ($p_{T}$) particles in the final state. Hence, such process is categorized as a single parton scattering, however other interactions occur besides the hard scattering, namely (i) initial state radiation, (ii) final state radiation, and (iii) interactions of beam remnants.

On the other hand, the high density of partons inside the colliding hadrons increase the probability of secondary interactions. Particles with low $p_{T}$ are observed at each hadronic collision, which are produced aside the high-$p_{T}$ particles in the event. This set of secondary interactions characterizes the MPI, and considering the secondary ones -- apart of the hard scattering and the beam remnants -- they are known as underlying events (UE). 

One of the main goals is to study this phenomenon to improve the models employed in event generators. Since it is not possible to disentangle the contribution from the hard scattering to the MPI in an event-by-event basis \cite{Aaboud:2017fwp}, one should look for observables sensitive to UE. To this end, both ATLAS and CMS Collaborations have investigated UE in the data obtained in proton-proton collisions at 13 TeV. A CMS analysis \cite{Sirunyan:2017vio} has focused on the event with inclusive $Z\to\mu^{+}\mu^{-}$ production to select high-$p_{T}$ particles as a reference for the UE. In a more broad study, the ATLAS Collaboration investigated UE observables in the production of charged particles \cite{Aaboud:2017fwp}. 
The observables for UE are divided into $\Delta\phi$ regions with respect to the particle with the highest $p_{T}$. The region of $|\Delta\phi|<$~60$^{\circ}$ around the leading particle is named {\it{towards}}, while the region of 60$^{\circ}$~$<|\Delta\phi|<$~120$^{\circ}$ is defined as {\it{transverse}}, and the region with $|\Delta\phi|>$~120$^{\circ}$ as {\it{away}}.

\begin{figure}[t]
\centering
%\begin{subfigure}[c]{0.5\textwidth}
\includegraphics[width=0.5\textwidth]{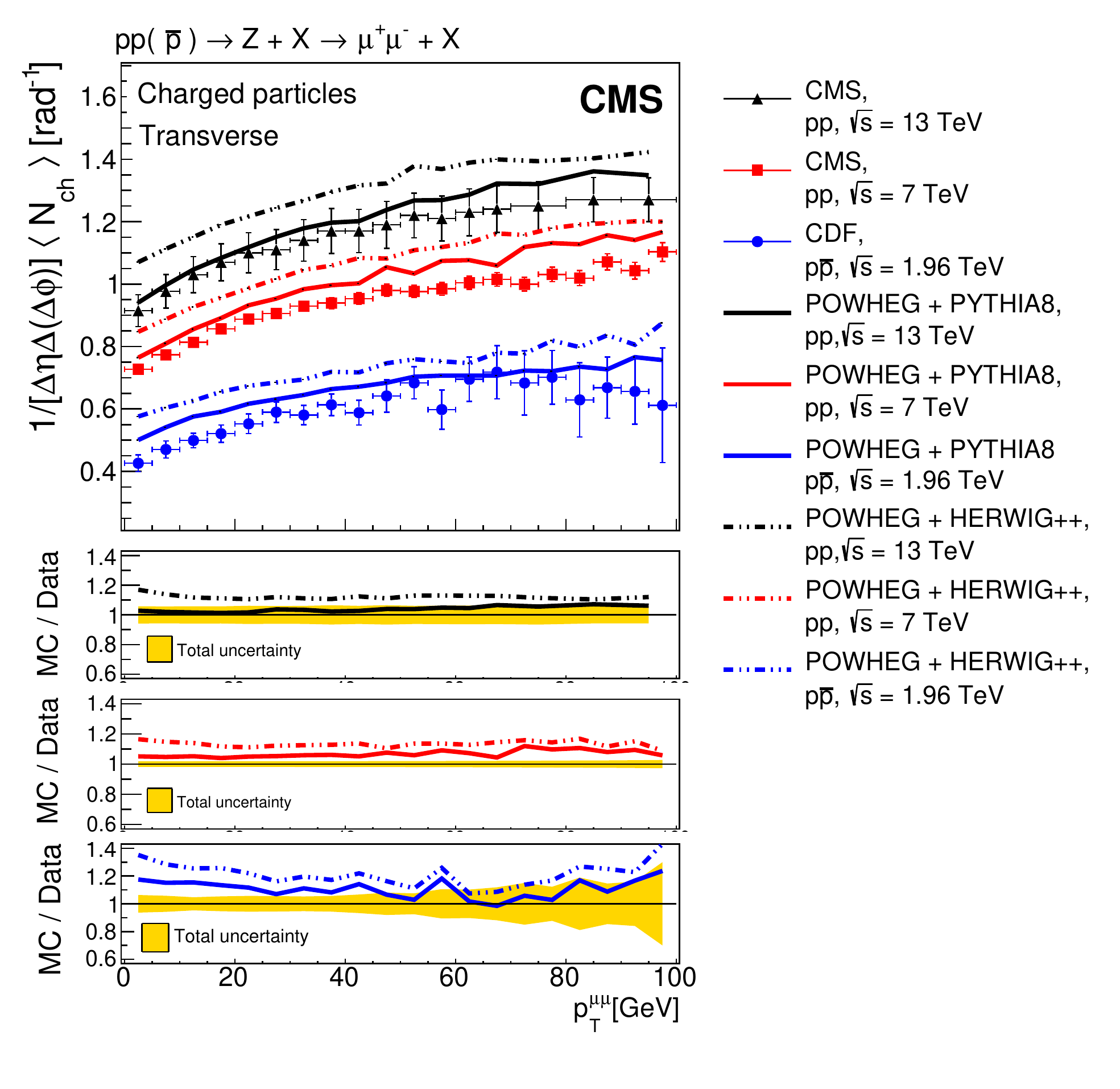}
%\end{subfigure}
%\begin{subfigure}[c]{0.35\textwidth}
\includegraphics[width=0.35\textwidth]{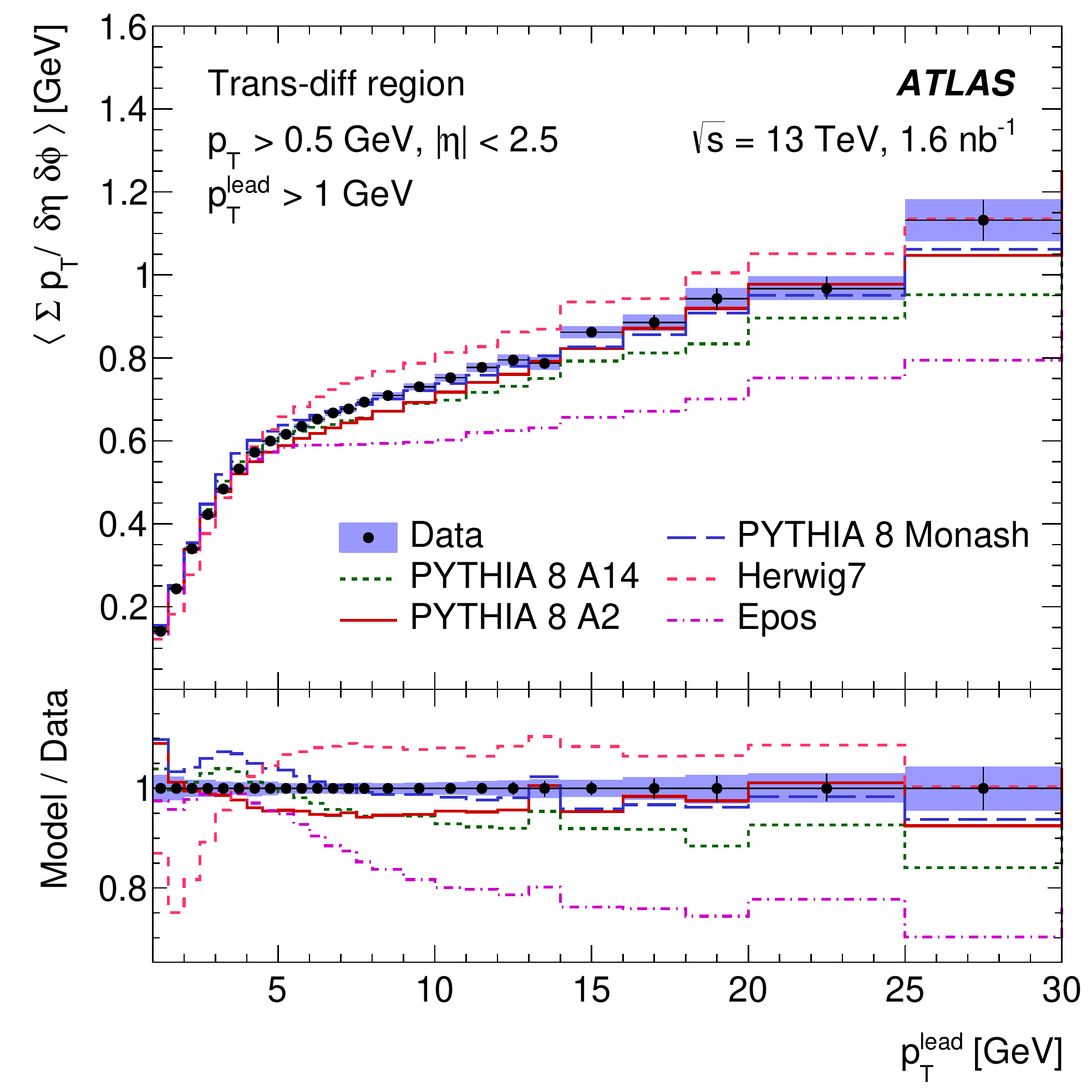}
%\end{subfigure}
\caption{\label{ue}(top) Particle density measured in $Z$ events in the transverse region as function of $p^{\mu\mu}_{T}$ \cite{Sirunyan:2017vio}. (bottom) Mean densities of charged-particle $\sum p_{T}$ in terms of the leading charged-particle $p_{T}$ in the transverse azimuthal region \cite{Aaboud:2017fwp}.}
\end{figure}

The results obtained by both experiments show that the MC modelling of UE is accurate to a 5\% level. The results reported by the CMS Collaboration show a comparison between the data recorded with the CDF experiment at the Tevatron ($\sqrt{s}$~=~1.96~TeV) and the data obtained at the LHC ($\sqrt{s}$~=~7~and~13~TeV). Figure~\ref{ue}(top) shows the comparison of these data sets with the MC predictions, showing that the results obtained with {\tt{POWHEG+PYTHIA8}} give a better description of the data than {\tt{POWHEG+HERWIG++}}. This comparison focus on the transverse region, which is more sensitive to the UE.

The results obtained by ATLAS are shown in Fig.~\ref{ue}(bottom), where the average charged particle multiplicity is present in terms of the transverse momentum of the leading particle. It is seen that the predictions obtained with {\tt{EPOS MC}} generator are in best agreement with the 13 TeV data in the transverse region in low $p_{T}$ regime which is more sensitive to the hard process. Although ATLAS has seen a good description of the minimum-bias data, the {\tt EPOS} event generator presents discrepant features with increasing $p_{T}^{\textrm{lead}}$, pointing to inconsistencies of the modelling of multiple interactions in $pp$ collisions at the LHC.

%%%%%%%%%%%%%%%%%%%%%%%%%%%%%%%%%%%%%%%%%%%%%%%%%%%%%%%%%%%%%%%%%%

\subsection{Double Parton Scattering}
\label{sec:dps}

Apart of the UE, there is a probability that another interaction occurs to produce a second hard scattering, resulting in more high-$p_{T}$ particles in the event \cite{Sjostrand:1987su}. This is the case of the double parton scattering (DPS), accounted by a effective cross section as follows:
\begin{eqnarray}
\sigma^{\textrm{DPS}}_{\textrm{AB}} = \frac{m}{2}\frac{\sigma_{\textrm{A}}\sigma_{\textrm{B}}}{\sigma_{\textrm{eff}}},
\end{eqnarray}
where $m=1$ is set for different pair production, $m=2$ for identical pairs, $\sigma_{A,B}$ is the $2\to 2$ cross section, and $\sigma_{\textrm{eff}}$ the effective cross section. This effective approach neglects the correlation between the partons in the hadrons.

The CMS experiment investigated the DPS signal using a multivariate model for the same-sign $WW$ production \cite{Sirunyan:2017hlu}. As a result, the CMS has found no excess in the data, being able to set an upper limit on the same-sign $WW$ cross section of 0.32~pb with 95\% of confidence level (CL) and a lower limit on the effective cross section of 12.2~mb with 95\% of CL.

Looking into the 4-jet production, the ATLAS Collaboration performed a similar investigation using Artificial Neural Networks \cite{Aaboud:2016dea}. The goal was to determine the fraction of DPS and to estimate $\sigma_{\textrm{eff}}$. The ATLAS experiment has found the following results:
\begin{eqnarray}
f_{\textrm{DPS}} &=& 0.092^{+0.005}_{-0.011} \textrm{ (stat.) } ^{+0.033}_{-0.037} \textrm{ (syst.) } \\
\sigma_{\textrm{eff}} &=& 14.9^{+1.2}_{-1.0} \textrm{ (stat.) } ^{+5.1}_{-3.8} \textrm{ (syst.) } \textrm{mb},
\end{eqnarray}
being $\sigma_{\textrm{eff}}$ consistent within uncertainties with previous results.

%%%%%%%%%%%%%%%%%%%%%%%%%%%%%%%%%%%%%%%%%%%%%%%%%%%%%%%%%%%%%%%%%%

\section{proton-Lead and Lead-Lead collisions}
\label{sec:pAAA}

%%%%%%%%%%%%%%%%%%%%%%%%%%%%%%%%%%%%%%%%%%%%%%%%%%%%%%%%%%%%%%%%%%

\subsection{Jet quenching}
\label{sec:quench}

In the collision of large systems, like heavy ions, the partons act like a deconfined colour charge system during a short-time scale, the so-called Quark Gluon Plasma (QGP). Given that the QGP is a hot medium, particles and jets produced in the collision may loose energy while crossing this medium, affecting their final-state kinematics. In particular, the energy lost by jets produced in the event reveals as an energy suppression: the jet quenching phenomenon. This effect was first observed at the Relativistic Hadron Ion Collider (RHIC) \cite{Adcox:2004mh,Adams:2005dq} and the ATLAS Collaboration has observed such phenomenon in PbPb collisions at 2.76~TeV \cite{Aad:2010bu}. The evidences suggested an imbalance in the $p_{T}$ distribution of jets, in agreement with the hypothesis of medium-induced energy loss.

\begin{figure}[htbp]
\centering
\includegraphics[width=0.5\textwidth]{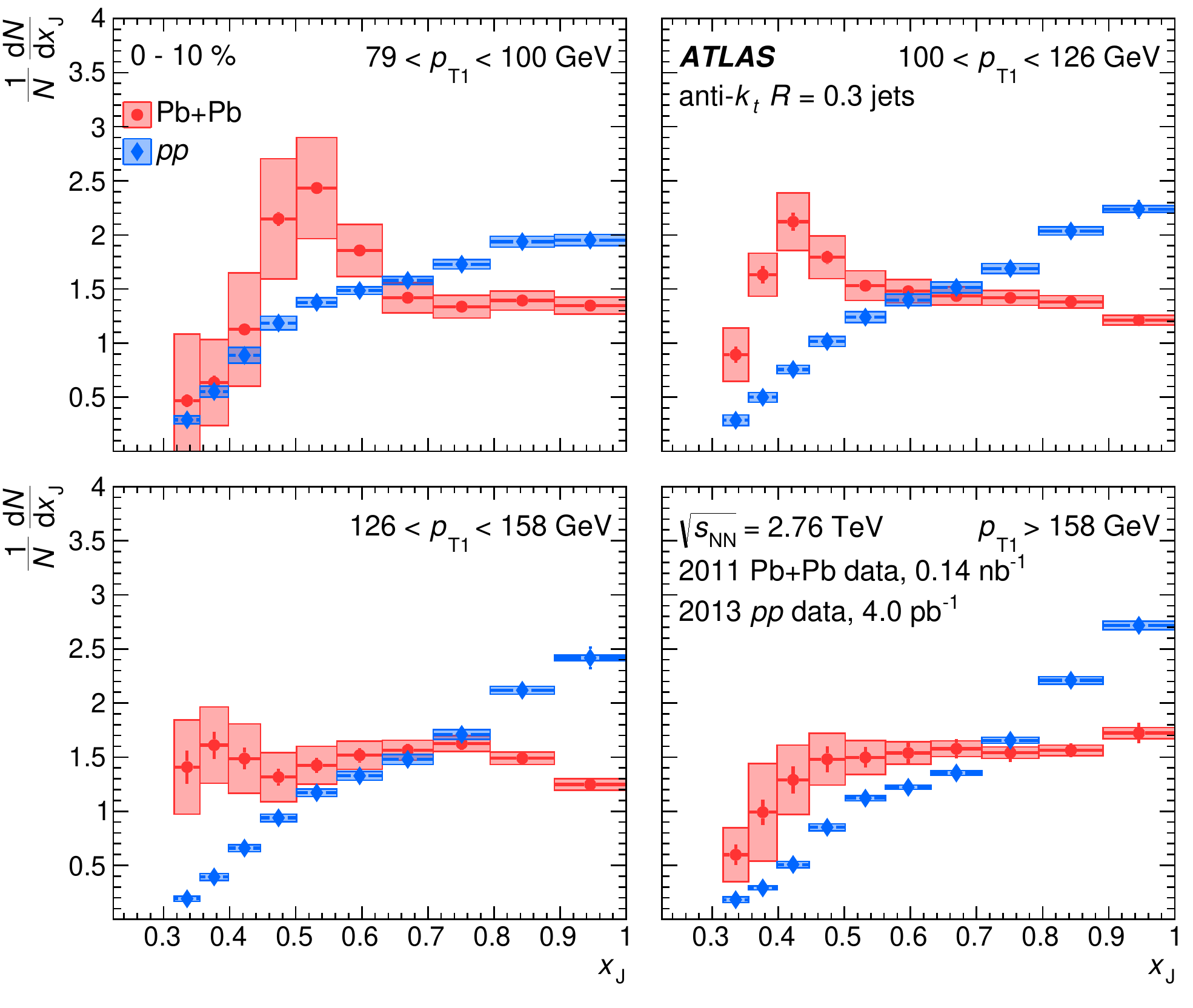}
\caption{\label{jetq} The dijet yield normalised by the total number of jet pairs in a given $p_{T1}$ interval in terms of $x_{J}$ \cite{Aaboud:2017eww}.}
\end{figure}

New ATLAS measurements present a deeper insight on this effect by looking at the $p_{T}$ correlations in distinct centralities in PbPb collisions \cite{Aaboud:2017eww}. These distributions are compared to those obtained in $pp$ collisions recorded in 2013 summing up 4/pb of data. In this analysis, a two-dimensional distribution in terms of the leading and sub-leading jets ($p_{T1}$,$p_{T2}$) are used to account for experimental effects. The Fig.~\ref{jetq} shows the fraction $x_{J} = p_{T2}/p_{T1}$ distribution in PbPb and $pp$ collisions at 2.76~TeV at centralities of 0\%--10\%. It is visible the effect of $p_{T}$ imbalance in PbPb collisions in comparison to the $pp$ case, results consistent with medium-induced energy loss expected in PbPb collisions due to the QGP.

%%%%%%%%%%%%%%%%%%%%%%%%%%%%%%%%%%%%%%%%%%%%%%%%%%%%%%%%%%%%%%%%%%

\subsection{Azimuthal correlations}
\label{sec:azimuth}

Considering that the hot medium may interfere in the kinematics of produced particles, these effects can be investigated using azimuthal distributions. In case that the hot medium is indeed present, the particle angular distribution is affected, giving rise to a spatial asymmetry when the collisions are not central \cite{Aaboud:2017acw}.

To study such asymmetries, the ATLAS Collaboration has employed the cumulant method to investigate azimuthal anisotropy in $pp$, $p$Pb, and low-multiplicity PbPb collisions at the LHC \cite{Aaboud:2017acw}. As a result, an increasing of the particle production is observed as a global long-range phenomenon that covers a wide range in pseudorapidity. The description of this effect is made by means of Fourier harmonics of the form $v_{n} = \langle \cos[n(\phi - \Phi_{R})] \rangle$, being $\phi$ the azimuthal angle and $\Phi_{R}$ the azimuthal angle of the reaction plane \cite{Poskanzer:1998yz}.

\begin{figure}[b]
\centering
\includegraphics[width=0.5\textwidth]{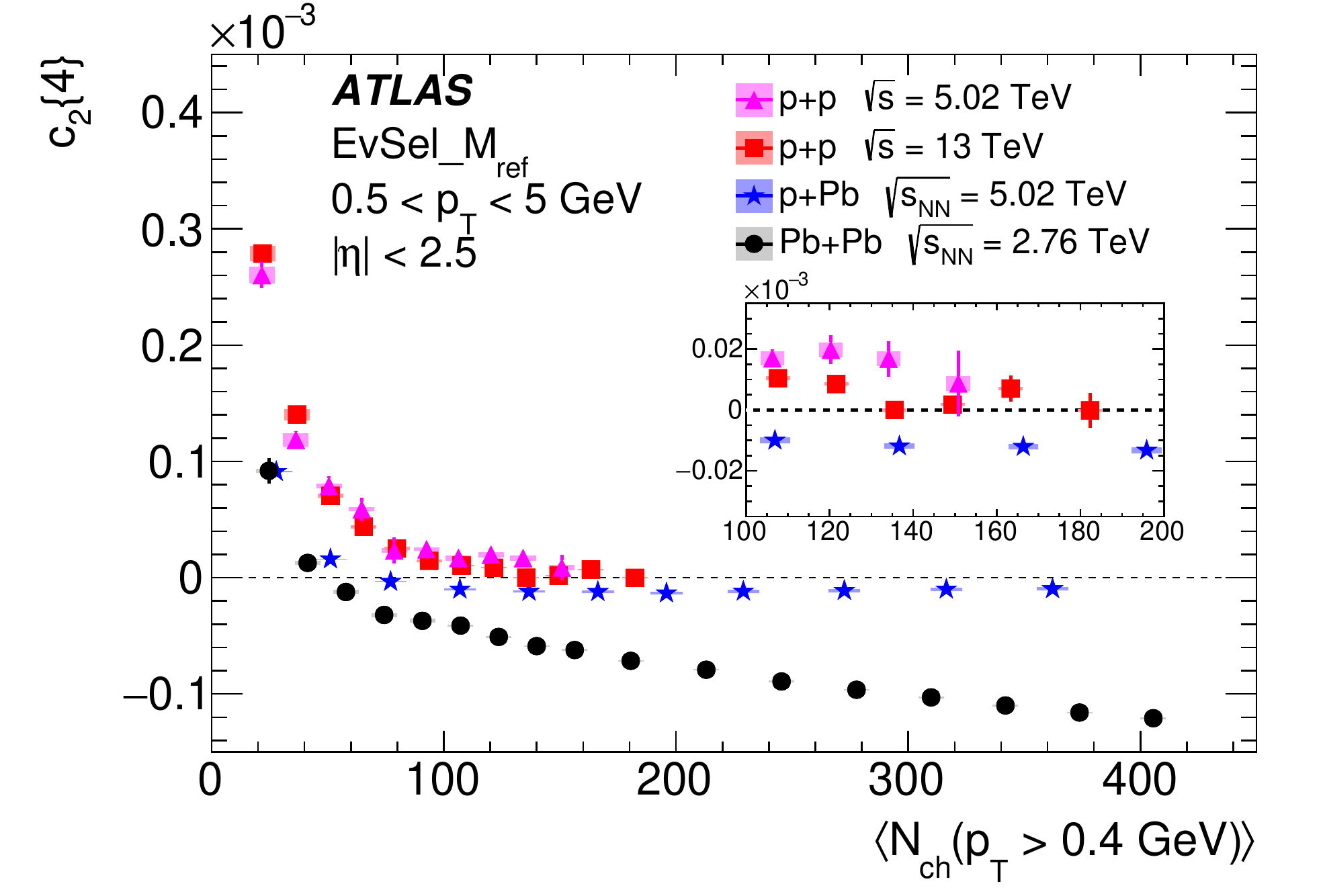}
\caption{\label{azcorr} The second-order cumulant $c_{2}\{4\}$ from four-particle correlations as function of the mean number of charged particle with $p_{T}>$~0.4~GeV for $pp$, $p$Pb, and PbPb collisions \cite{Aaboud:2017acw}.}
\end{figure}

The ATLAS experiment has analysed $pp$ datasets from 2015 at 5.02~TeV and 13~TeV, a $p$Pb dataset from 2013 at 5.02~TeV, and a PbPb dataset from 2014 at 2.76~TeV. These data are used to evaluate the long-range two-particle azimuthal correlation, extracting the Fourier harmonics $v_{n}$, $n=2,3,4$. Taking the different colliding systems, it is expected that the effect of azimuthal correlation in $pp$ collisions is much weaker than in $p$Pb and PbPb ones. Besides, the $p$Pb system is used to study the cold nuclear matter (CNM), which is expected to be weaker than in PbPb collisions. Consequently, the ATLAS experiment has found that the four-particle correlation function in $pp$ collisions does not provide an unbiased measurement of the $v_{2}$ coefficient by the cumulant method and not show an energy or multiplicity dependence. Besides, Figure~\ref{azcorr} shows the second-order coefficient in terms of the mean charged-particle multiplicity, which shows an ordering of $|c_{2}\{4\}|_{\mathrm{p+Pb}} < |c_{2}\{4\}|_{\mathrm{Pb+Pb}}$ for $N_{\textrm{ch}}(p_{T}>$~0.4~GeV)~$>100$. In summary, the cumulant and the second-order harmonic increase in $p$Pb and PbPb collisions for higher multiplicities. 

\begin{figure*}
\centering
\includegraphics[width=\textwidth]{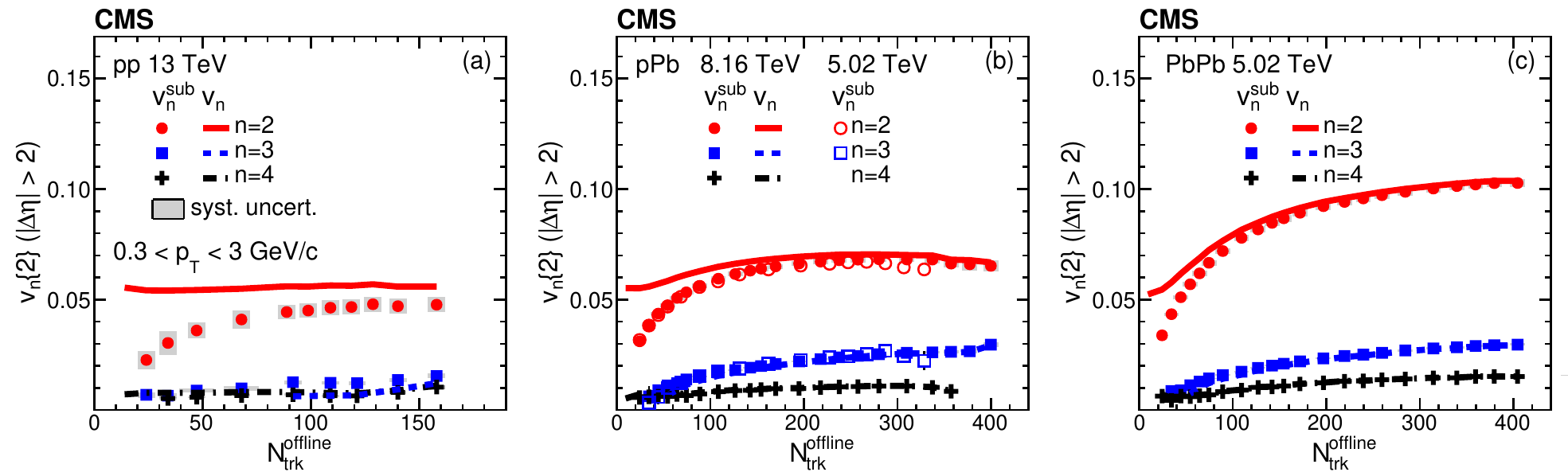}
\caption{\label{cms-ridge} The $v_{2,3}$ (from \cite{Khachatryan:2016txc}) and $v_{4}$ coefficients for long-range two-particle correlations in terms of the charged-particle multiplicity, $N^{\textrm{offline}}_{\textrm{trk}}$, for $pp$, $p$Pb, and PbPb collisions The solid line represents the results before the low-multiplicity subtraction \cite{Sirunyan:2017uyl}.}
\end{figure*}

%%%%%%%%%%%%%%%%%%%%%%%%%%%%%%%%%%%%%%%%%%%%%%%%%%%%%%%%%%%%%%%%%%

\subsection{The Ridge}
\label{sec:ridge}

Another relevant effect that can be studied with multiparticle correlations is the {\it{ridge}} effect: the two-particle correlation function shows that there is a collimated emission of pairs in a small $\Delta\phi$ range and a wide range in $\Delta\eta$. The experimental signature of the ridge is studied by a Fourier decomposition of the correlation function as:
\begin{eqnarray}
C(\Delta\phi) \sim 1 + 2\sum_{n} v^{2}_{n}\cos(n\Delta\phi), 
\end{eqnarray}
where $v_{n}$ is the single-particle anisotropy harmonic coefficient and carry information about the collective behaviour.

The CMS experiment has performed the measurement of the anisotropy coefficient $v_{4}$ at high precision in different colliding energies for $pp$, $p$Pb, and PbPb collisions, especially with the latest data of $p$Pb collisions at 8.16~TeV \cite{Sirunyan:2017uyl}. Furthermore, a measurement of the $v_{n}$ coefficients are extracted via long-range, $\Delta\eta >2$, two-particle correlations in terms of the charged particle multiplicity. Figure~\ref{cms-ridge} shows the coefficient versus the multiplicity for $v_{2}$, $v_{3}$, and $v_{4}$ from the three datasets analysed. The solid lines represent the results before correcting for the low-multiplicity subtraction, which is more relevant at low multiplicities and plays a minor role at high multiplicities. Hence, the results obtained at 8.16~TeV are consistent with those obtained at 5.02~TeV. Besides, the four-particle cumulant method has show evidences that for a hydrodynamic flow of a strongly interacting medium, similar to what is observed in PbPb systems.

Using an improve cumulant method, the ATLAS Collaboration is interested in investigating whether the ridge effect results from a collective flow or other correlations by a few particles \cite{Aaboud:2017blb}. This method explores the correlation between subevents separated in $\eta$ to reduce the non-flow correlations \cite{Jia:2017hbm}, which is validated using the {\tt{PYTHIA8}} event generator \cite{Sjostrand:2007gs}. As a result, the ATLAS Collaboration has found that the previous cumulant method used in $pp$ collisions may be contaminated with non-flow correlations, and the new (improved) method provides results more consistent with long-range collective flow correlations.

%%%%%%%%%%%%%%%%%%%%%%%%%%%%%%%%%%%%%%%%%%%%%%%%%%%%%%%%%%%%%%%%%%

\subsection{Vector meson production}
\label{sec:meson}

In view of direct particle production, the LHCb experiment has reported the measurement of different vector mesons in $p$Pb collisions. At 5~TeV the LHCb studies the prompt $D^{0}$ meson production, and at 8.16~TeV the $J/\psi$ meson production, both meant to investigate the CNM effect in nuclear collisions \cite{Aaij:2017cqq}. The $p$Pb collisions provide the best system to study the CNM effects, which has been done in fixed-target experiments and at RHIC and the LHC \cite{Andronic:2015wma}. Moreover, such measurements can provide constraints to the low-$x$ gluons in nuclei, improving the data for nuclear parton distribution functions.

\begin{figure}[b]
\centering
\includegraphics[width=0.4\textwidth]{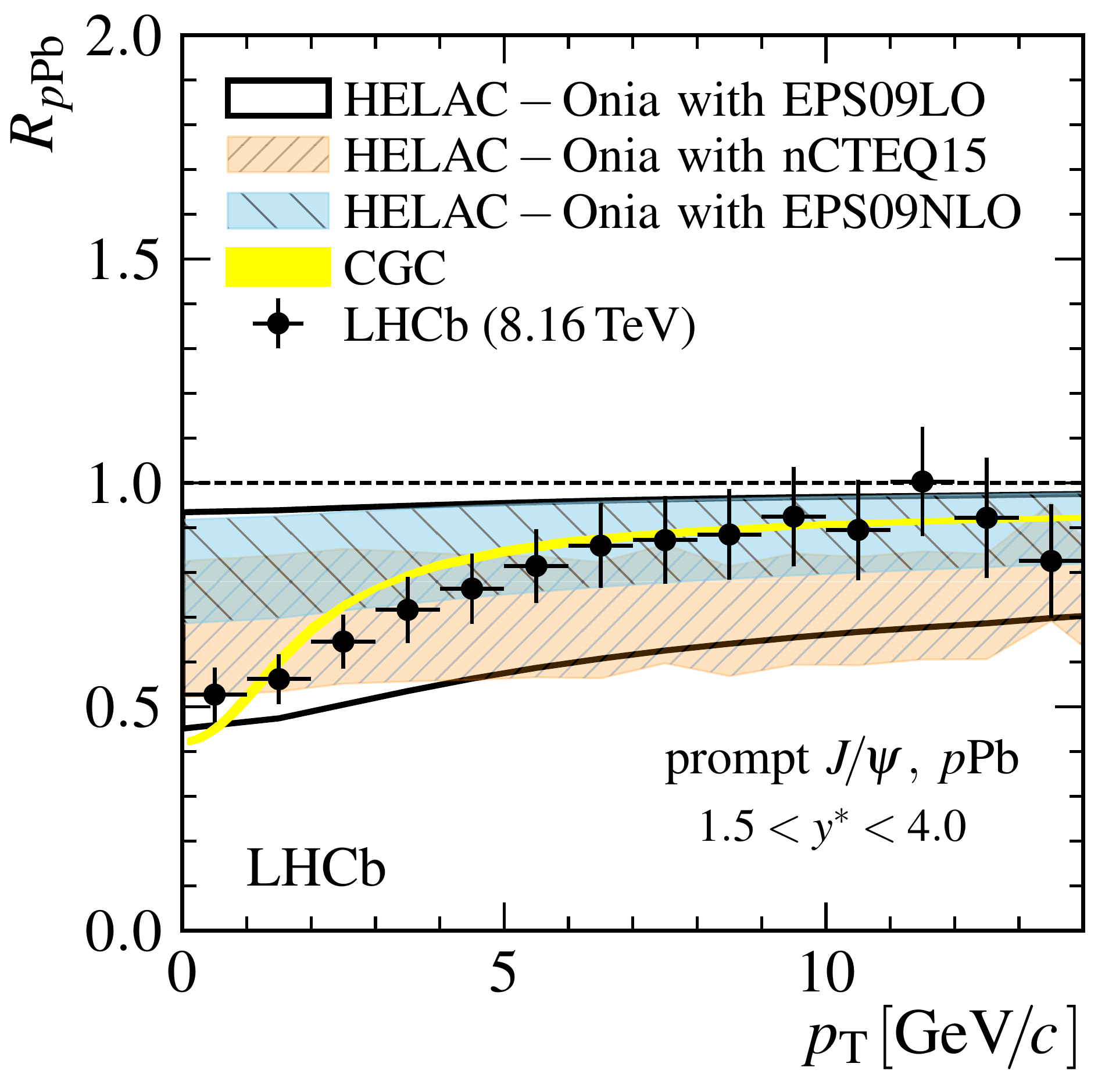}
\caption{\label{meson} Nuclear modification factor as function of $p_{T}$ for prompt $J/\psi$ production in $p$Pb collisions, integrated over $y^{*}$. The black circles are the data and the coloured bands are predictions from different models \cite{Aaij:2017cqq}.}
\end{figure}

One of the key observables in nuclear collisions is the nuclear modification factor,
\begin{eqnarray}
R_{p\mathrm{Pb}} = \frac{1}{A}\frac{\dif^{\,2}\sigma_{p\textrm{Pb}}(p_{T},y^{*})/\dif p_{T}\dif y^{*}}{\dif^{\,2}\sigma_{pp}(p_{T},y^{*})/\dif p_{T}\dif y^{*}},
\end{eqnarray}
where $A=208$ and $y^{*}$ is the rapidity in the centre-of-mass frame of the colliding nucleons. Figure~\ref{meson} shows the LHCb data for the $J/\psi$ production and the predictions from different phenomenological models. It is noticeable that the colour glass condensate (CGC) framework does a good job in describing the data in the forward rapidity region. A similar plot is presented in \cite{Aaij:2017gcy}, where {\tt{EPOS09NLO}} does a better job at low masses than the CGC predictions for the $D^{0}$ production.

\begin{figure*}
\centering
\includegraphics[width=0.48\textwidth]{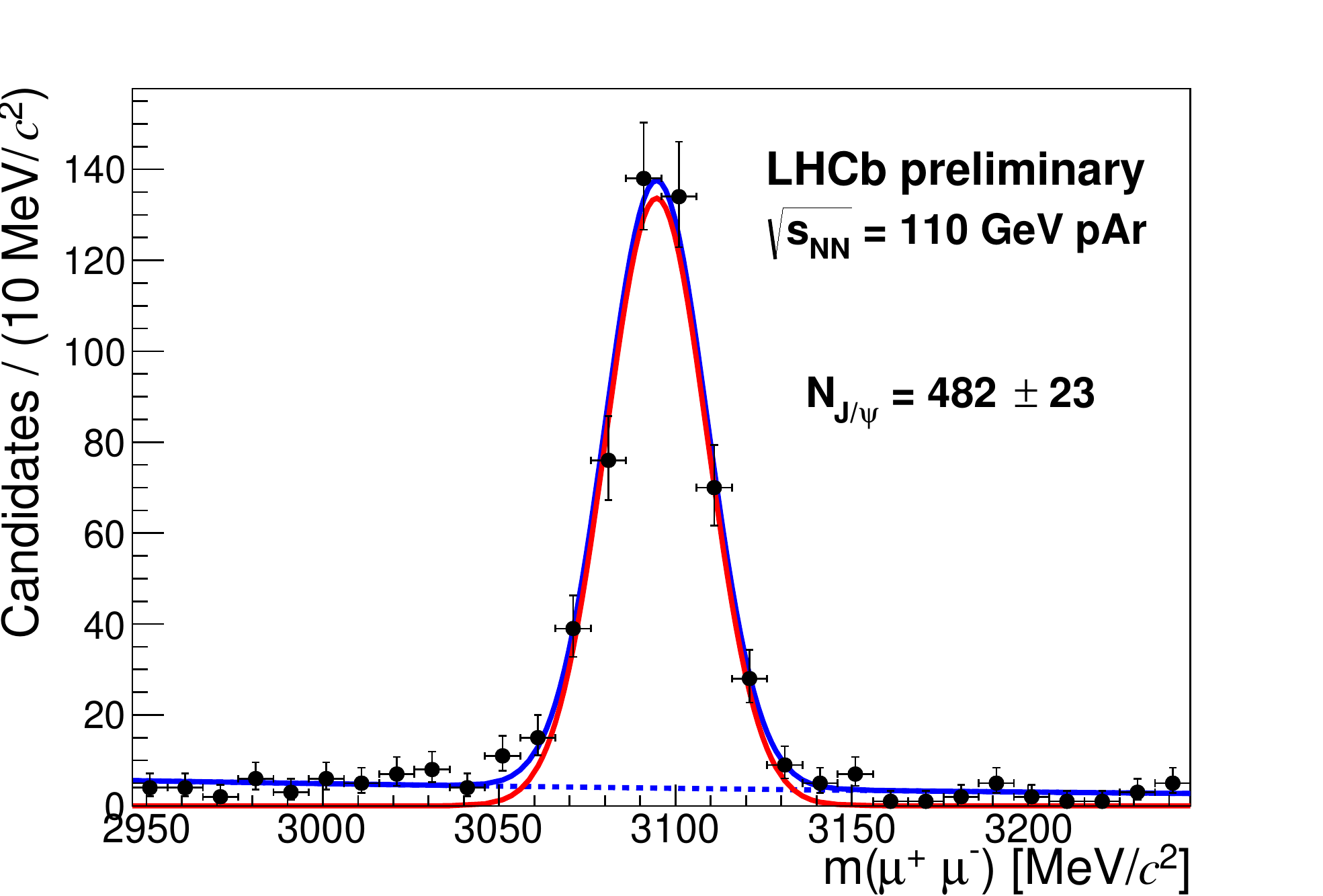}
\includegraphics[width=0.48\textwidth]{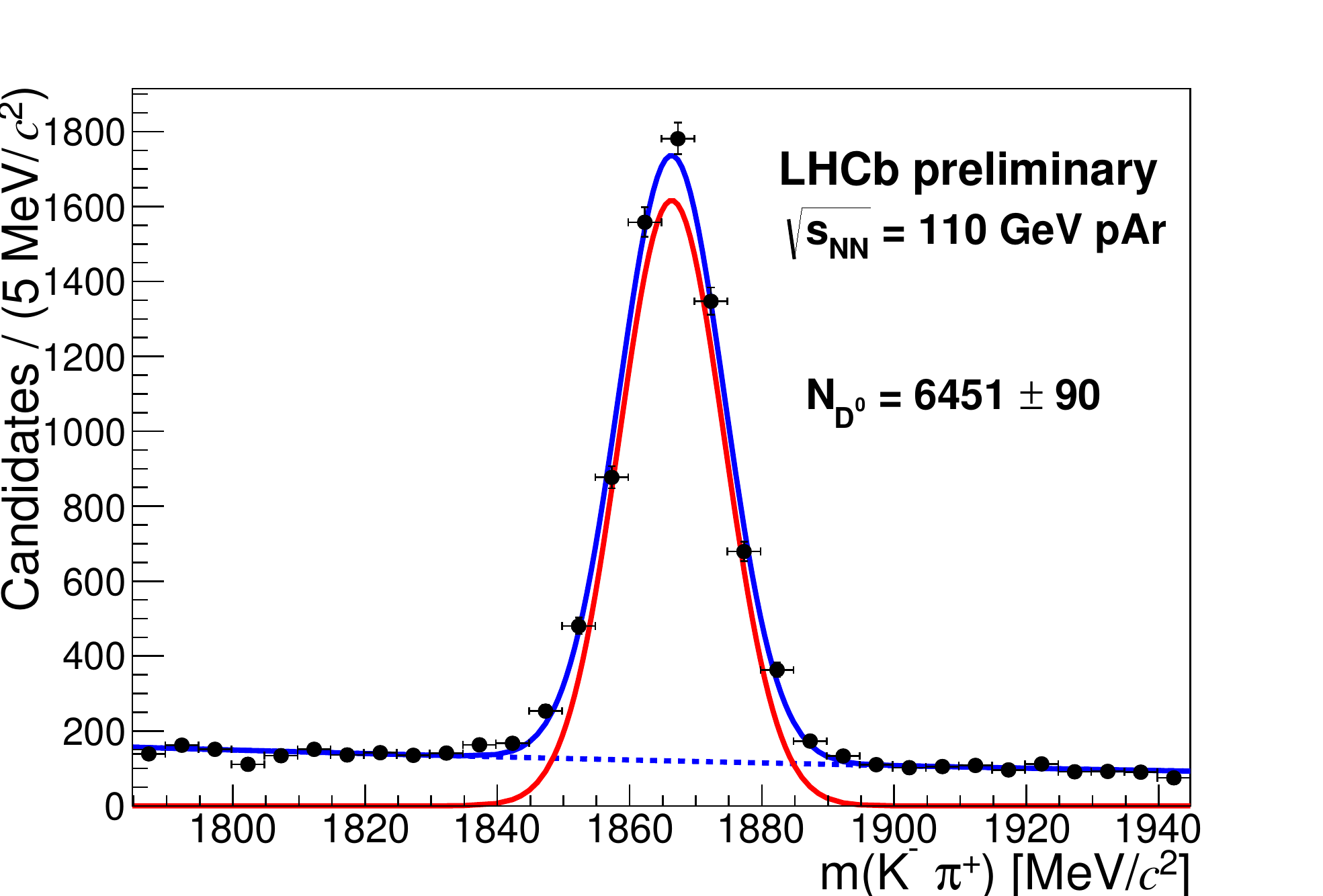}
\caption{\label{pAr} Invariant mass distribution for (left) $J/\psi\to\mu^{+}\mu^{-}$ and (right) $D^{0}\to K^{-}\pi^{+}$. The black circles are the data and the solid lines represent the fit: red line is the signal, blue-dashed line the continuum background, and the blue line the signal+background \cite{LHCb:2017qap}.}
\end{figure*}

%%%%%%%%%%%%%%%%%%%%%%%%%%%%%%%%%%%%%%%%%%%%%%%%%%%%%%%%%%%%%%%%%%

\subsection{Fixed-target results}
\label{sec:fixed}

The LHCb Collaboration also provides interesting results using the System for Measuring Overlap with Gas (SMOG), where very-low pressure gas is injected to perform measurements in fixed-target mode. In employing Argon gas, the LHCb experiment is able to perform fixed-target $p$Ar collisions at 110.4~GeV (in contrast to typical TeV $pp$ collisions) \cite{LHCb:2017qap}. These low-energy collisions that produce charmonium states are important to understand the charmonium suppression in heavy-ion collisions, being such phenomenon connected to the formation of the QGP in nuclear collisions, where the charmonium production is used as a thermometer of the system.

Using the SMOG system, the LHCb experiment measured $J/\psi\to\mu^{+}\mu^{-}$ and $D^{0}\to K^{-}\pi^{+}$, which are important to constraint the intrinsic charm contributions \cite{Pumplin:2007wg}. Figure~\ref{pAr} shows the invariant mass distributions with the corresponding resonances, resulting in a yield of 500 $J/\psi\to\mu^{+}\mu^{-}$ and 6 500 $D^{0}\to K^{-}\pi^{+}$ and showing the feasibility of a fixed-target experiment at the LHC.

The LHCb experiment also explored $p$He collisions with the SMOG system to produce antiprotons at 110.4~GeV \cite{LHCb:2017tqz}. This measurement is of importance to constraints the antimatter content in cosmic rays (CR) and its propagation in the interstellar medium. Moreover, such investigation can also provide evidences for an indirect probe for Dark Matter. A key observable is the $\bar{p}/p$ ratio, which is expected to be of the order of 10$^{-4}$. Recent experimental efforts with the PAMELA \cite{Adriani:2012paa} and AMS-02 \cite{Aguilar:2016kjl} experiments have provided new grounds to improve the measurement for the $\bar{p}/p$ ratio, reporting relevant results for the CR content.

The $\bar{p}$ production measured at the LHCb covers the $\bar{p}$ momentum range from 12~to~110~GeV/$c$, both prompt $\bar{p}$ production and from decay of resonances. The LHCb measured more than 1 million antiprotons with a measured production cross section of:
\begin{eqnarray}
\sigma^{\mathrm{LHCb}}_{\mathrm{inel}} (p\mathrm{He},\sqrt{s_{\mathrm{NN}}} = 100\textrm{ GeV}) = (140\pm 10)\textrm{ mb}.
\end{eqnarray}

\begin{figure*}[t]
\centering
\includegraphics[width=0.6\textwidth]{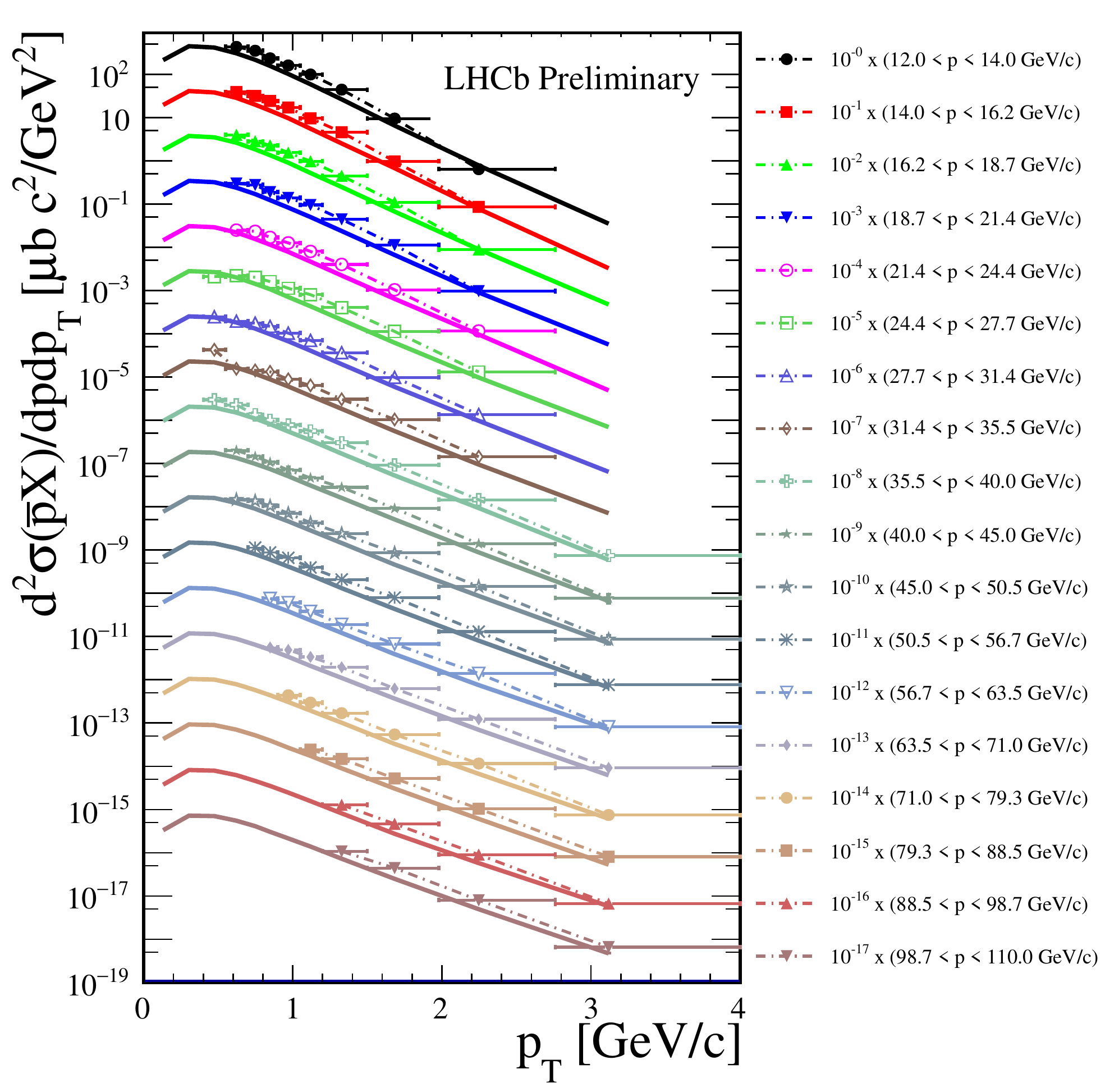}
\caption{\label{antip} Double differential $\bar{p}$ cross section in terms of $p_{T}$ in 18 momentum bins, all scaled by a factor of 0.1. The lines show the predictions from {\tt{EPOS LHC}}, scaled with the same factor as data \cite{LHCb:2017tqz}.}
\end{figure*}

Also, the LHCb Collaboration provides de double differential cross section in terms of $p_{T}$ for 18 momentum bins, shown in Fig.~\ref{antip}. The data is compared to the predictions obtained with the {\tt{EPOC LHC}} event generator \cite{Pierog:2013ria}, presenting an overall good agreement, although the absolute production rate is off by a factor of 1.5. The ratio of the measured cross section and the predicted one is:
\begin{eqnarray}
\frac{\sigma^{\mathrm{LHCb}}_{\mathrm{inel}}}{\sigma^{\mathrm{EPOS}}_{\mathrm{inel}}} = 1.19\pm 0.08,
\end{eqnarray}
where the event generator predicts a smaller multiplicity per inelastic collision as seen in data.

%%%%%%%%%%%%%%%%%%%%%%%%%%%%%%%%%%%%%%%%%%%%%%%%%%%%%%%%%%%%%%%%%%

\section{Summary} 
\label{sec:sum}

A set of recent results from the LHC experiments is reported for $pp$, $p$Pb, and PbPb collisions, performing more precise and detailed investigations on various topics on high-energy physics. In $pp$ collisions, the ATLAS and CMS Collaborations have studied strong interactions in UE and DPS, providing results to constraint phenomenological models. In nuclear collisions, various effects have been studied, from azimuthal correlations to antiproton production. These results show the capabilities of the LHC experiments to perform measurements in a complex environment and to provide deep insight on key observables to enrich our understanding on many aspects of high-energy physics.

%%%%%%%%%%%%%%%%%%%%%%%%%%%%%%%%%%%%%%%%%%%%%%%%%%%%%%%%%%%%%%%%%%

\begin{acknowledgments}
GGS thanks the partially support by CNPq, CAPES, and FAPERGS, Brazil.
\end{acknowledgments}

%%%%%%%%%%%%%%%%%%%%%%%%%%%%%%%%%%%%%%%%%%%%%%%%%%%%%%%%%%%%%%%%%%
\bibliographystyle{apsrev4-1}

%%%%%%%%%%%%%%%%%%%%%%%%%%%%%%%%%%%%%%%%%%%%%%%%%%%%%%%%%%%%%%%%%%

\end{document}